\documentclass{PoS}

\usepackage{graphicx}
\usepackage{amssymb}
\usepackage{amsmath}

 \newcommand{\preprint}{
   \begin{picture}(0,0)
     \put(390,170){{\rm\normalsize ADP-10-19/T715}}
   \end{picture}}

\title{\preprint%
 Excited States of the Nucleon in 2+1 Flavor QCD}

\ShortTitle{}

\author{M. S. Mahbub $^{a\,b}$, \, W. Kamleh $^{a}$, \, \speaker{D.
    B. Leinweber} $^{a}$, P. J. Moran $^{a}$, A. G. Williams
  $^{a}$ \\
       \llap{$^a$} Special Research Centre for the Subatomic Structure
       of  Matter, Adelaide, South Australia 5005, Australia, and
       Department of Physics, University of Adelaide, South  
       Australia 5005, Australia.\\
       \llap{$^b$} Department of Physics, Rajshahi University,
       Rajshahi 6205, Bangladesh.\\ 

       E-mail: \email {md.mahbub@adelaide.edu.au, \\
                       waseem.kamleh@adelaide.edu.au, \\ 
                       derek.leinweber@adelaide.edu.au, \\ 
                       peter.moran@alumni.adelaide.edu.au, \\
                       anthony.williams@adelaide.edu.au}}

\abstract{Recent developments on the determination of the spin-1/2 spectrum of
the nucleon in full QCD are presented.  Our focus is on the PACS-CS
2+1 flavor configurations made available through the ILDG.  Using
correlation matrix techniques, in which a wide variety of
gauge-invariant Gaussian-smeared fermion-propagator sources and sinks
are considered, excited states are determined.  We consider several
correlation matrices of various sizes, each constructed with a
different set of basis interpolators, in order to demonstrate the
invariance of the eigenstates on the basis choice.  Of particular
interest is the approach to the elusive Roper resonance and we
report preliminary results in full QCD.
}

\FullConference{The XXVIII International Symposium on Lattice Field Theory, Lattice2010\\
		June 14-19, 2010\\
		Villasimius, Italy}

\begin{document}

\section{Introduction}
The first positive-parity excited state of the nucleon, known as the
Roper resonance, $N{\frac{1}{2}}^{+}$(1440 MeV) ${\rm P}_{11}$, has
presented a long-standing puzzle
since its discovery in the 1960's due to its lower mass compared to the adjacent
negative parity, $N{\frac{1}{2}}^{-}$(1535 MeV) ${\rm S}_{11}$, state.    
In constituent quark models with harmonic oscillator
potentials, the lowest-lying odd-parity state naturally occurs below
the ${\rm P}_{11}$ state~\cite{Isgur:1977ef,Isgur:1978wd}. In nature the Roper
resonance is almost 100 MeV below the ${\rm S}_{11}$ state. 

Lattice QCD is very successful in computing many properties of hadrons
from first principles. In particular, the ground state of the hadron
spectrum is a well understood problem. However, the
excited states still provide a significant challenge. The
first analysis of the positive parity excitation of the nucleon
was performed in Ref.~\cite{Leinweber:1994nm} using Wilson fermions
and an operator product expansion spectral ansatz. 
Since then several attempts have been made to address these issues in
the  lattice framework.

The ``variational method''~\cite{Michael:1985ne,Luscher:1990ck} is one of
 the  state-of-the-art approaches to hadron  spectroscopy, which is
based on a correlation matrix analysis. The identification of the
Roper state with this method wasn't successful until recently. 
In Refs.~\cite{Mahbub:2009aa,Mahbub:2010jz} a low-lying Roper
state has been identified with this approach by employing a diverse
range of smeared-smeared correlation functions. Also, with this method, a
physical level ordering between the Roper and $N{\frac{1}{2}}^{-}$ ground state
is observed in Ref.~\cite{Mahbub:2010me}.

 Recent developments of algorithms and computational power have enabled
us to explore this physics in full QCD. Some recent full QCD results has
been presented in Refs.~\cite{Bulava:2009jb,Bulava:2010yg,Engel:2010my}. In this
paper,  we present preliminary  results for the excited states of the nucleon
 using the PACS-CS 2+1 flavor configurations~\cite{Aoki:2008sm}. The
 results are presented from the variational method using various
 fermion source  and sink smearings to construct correlation matrices.

\section{Variational Method}
 \label{sec:variational_method}

The two point correlation function matrix for a $\vec{p} =0$ baryon
can be  written as
\begin{align}
G_{ij}^{\pm}(t) &= \sum_{\vec x}{\rm Tr}_{\rm sp}\{ \Gamma_{\pm}\langle\Omega\vert\chi_{i}(x)\bar\chi_{j}(0)\vert\Omega\rangle\}, \\
          &=\sum_{\alpha}\lambda_{i}^{\alpha}\bar\lambda_{j}^{\alpha}e^{-m_{\alpha}t},
\end{align}
where Dirac indices are implicit. Here, $\lambda_{i}^{\alpha}$ and
$\bar\lambda_{j}^{\alpha}$ are the couplings of interpolators $\chi_{i}$ and
$\bar\chi_{j}$ at the sink and source respectively and $\alpha$
enumerates the energy eigenstates with mass
$m_{\alpha}$. $\Gamma_{\pm}=\frac{1}{2}(\gamma_{0}\pm 1)$ projects the parity of the eigenstates.\\ 
 Since the only $t$ dependence comes from the exponential term, one
 can seek a linear superposition of interpolators,
 ${\bar\chi}_{j}u_{j}^{\alpha}$, such that,  
\begin{align}
G_{ij}(t_{0}+\triangle t)\, u_{j}^{\alpha} & = e^{-m_{\alpha}\triangle
  t}\, G_{ij}(t_{0})\, u_{j}^{\alpha},
\end{align}  
for sufficiently large $t_{0}$ and $t_{0}+\triangle t$. More detail can be found in Refs.~\cite{Melnitchouk:2002eg,Mahbub:2009nr,Blossier:2009kd}. Multiplying the above equation by $[G_{ij}(t_{0})]^{-1}$ from the left leads to an eigenvalue equation,
\begin{align}
[(G(t_{0}))^{-1}G(t_{0}+\triangle t)]_{ij}\, u^{\alpha}_{j} & = c^{\alpha}\, u^{\alpha}_{i},
 \label{eq:right_evalue_eq}
\end{align} 
where $c^{\alpha}=e^{-m_{\alpha}\triangle t}$ is the eigenvalue. Similar to Eq.~(\ref{eq:right_evalue_eq}), one can also solve the left eigenvalue equation to recover the $v^{\alpha}$ eigenvector,
\begin{align}
v^{\alpha}_{i}\, [G(t_{0}+\triangle t)(G(t_{0}))^{-1}]_{ij} & = c^{\alpha}v^{\alpha}_{j}.
\label{eq:left_evalue_eq}
\end{align} 
The vectors $u_{j}^{\alpha}$ and $v_{i}^{\alpha}$  diagonalize the correlation matrix at time $t_{0}$ and $t_{0}+\triangle t$ making the projected correlation matrix,

\begin{align}
v_{i}^{\alpha}G_{ij}^{\pm}(t)u_{j}^{\beta} & \propto \delta^{\alpha\beta}.
 \label{projected_cf} 
\end{align} 
The parity projected, eigenstate projected correlator, 
\begin{align}
 G^{\alpha}_{\pm}& \equiv v_{i}^{\alpha}G^{\pm}_{ij}(t)u_{j}^{\alpha} ,
 \label{projected_cf_final}
\end{align}
 is then analyzed using standard techniques to obtain the masses of
 the states, $\alpha$.

\section{Simulation Details}

The PACS-CS $2+1$ flavor dynamical-fermion
configurations~\cite{Aoki:2008sm} are used. The
non-perturbatively ${\mathcal{O}}(a)$-improved Wilson fermion
action and Iwasaki-gauge action~\cite{Iwasaki:1983ck} are employed. The
lattice volume is $32^{3}\times 64$, with $\beta=1.90$ providing
lattice spacing $a=0.0907$ fm and $c_{SW}=1.715$~\cite{Aoki:2005et}. Five
kappa values with degenerate up down quarks have been considered,
i.e.~$k_{ud}$ = 0.13700, 0.13727, 0.13754, 0.13770, 0.13781, with
$k_{s}=0.13640$. In this paper, we also consider the Sommer
scale~\cite{Sommer:1993ce}, with $a$ as described in
Ref.~\cite{Aoki:2008sm}.  The results are presented
for $k_{ud}\,=\, 0.13754,0.13770$, for which ensembles of 350
configurations are considered. As with the quenched
case~\cite{Mahbub:2009aa},  various levels  of gauge invariant Gaussian smearing
~\cite{Gusken:1989qx} are applied at the fermion source ($t=16$)
and at the sink. We consider 4, 9, 16, 25, 35,  50, 70, 100, 125,
200, 400, 800, 1600 sweeps, corresponding to rms radii in lattice
units of 1.20, 1.79, 2.37, 2.96, 3.50, 4.19, 4.95, 5.920,
6.63, 8.55, 12.67, 15.47, 16.00. The error analysis is performed using
the jackknife method, where the ${\chi^{2}}/{\rm{dof}}$ is obtained
via a  covariance matrix analysis. Our fitting method is discussed
extensively in Refs.~\cite{Mahbub:2010jz,Mahbub:2009nr}. 
      
The nucleon interpolator we consider here is the local scalar-diquark
interpolator ~\cite{Leinweber:1994nm,Leinweber:1990dv},
\begin{align}
\chi_1(x) &= \epsilon^{abc}(u^{Ta}(x)\, C{\gamma_5}\, d^b(x))\,
u^{c}(x).
\label{eqn:chi1_interpolator}
\end{align}

\begin{table}[!h]
    \begin{center}
   \caption{$4\times 4$ correlation matrix bases.} \vspace{0.5cm}
 \label{table:varieties_of_4x4_matrices}   
   \begin{tabular}{c|cccccccccc} 
        \hline
     Sweeps $\rightarrow$  & 16 & 25 & 35 & 50 & 70 & 100 & 125 & 200 & 400 & 800  \\
        \hline
   Basis No. $\downarrow$ & \multicolumn{10}{c}{Bases}  \\
        \hline 

1  & 16 & -  & 35 & -   & 70 & 100 & -   & -   & -   & - \\ 
2  & 16 & -  & 35 & -   & 70 & -   & 125 & -   & -   & - \\
3  & 16 & -  & 35 & -   & -  & 100 & -   & 200 & -   & - \\
4  & 16 & -  & 35 & -   & -  & 100 & -   & -   & 400 & - \\
5  & 16 & -  & -  & 50  & -  & 100 & 125 &  -  & -   & - \\
6  & 16 & -  & -  & 50  & -  & 100 & -   & 200 & -   & - \\
7  & 16 & -  & -  & 50  & -  & -   & 125 &  -  & -   & 800 \\
8  & -  & 25 & -  & 50  & -  & 100 & -   & 200 & -   & - \\
9  & -  & 25 & -  & 50  & -  & 100 & -   & -   & 400 & - \\
10 & -  & -  & 35 & -   & 70 & -   & 125 & -   & 400 & - \\ 

\hline 
 
 \end{tabular}
 \end{center}
\end{table}

\section{Results}

We consider several $4\times 4$ correlation matrices. Each matrix is
constructed with different sets of correlation functions, each set
element corresponding to 
a different number of sweeps of gauge invariant Gaussian smearing at the
source and sink of the $\chi_{1}\bar\chi_{1}$ correlators. This provides a
large basis of operators as described in
Table~\ref{table:varieties_of_4x4_matrices}, providing a wide range of
overlap  with energy eigenstates.

\begin{figure*}[!t]
  \begin{center}
 \includegraphics[height=0.98\textwidth,angle=90]{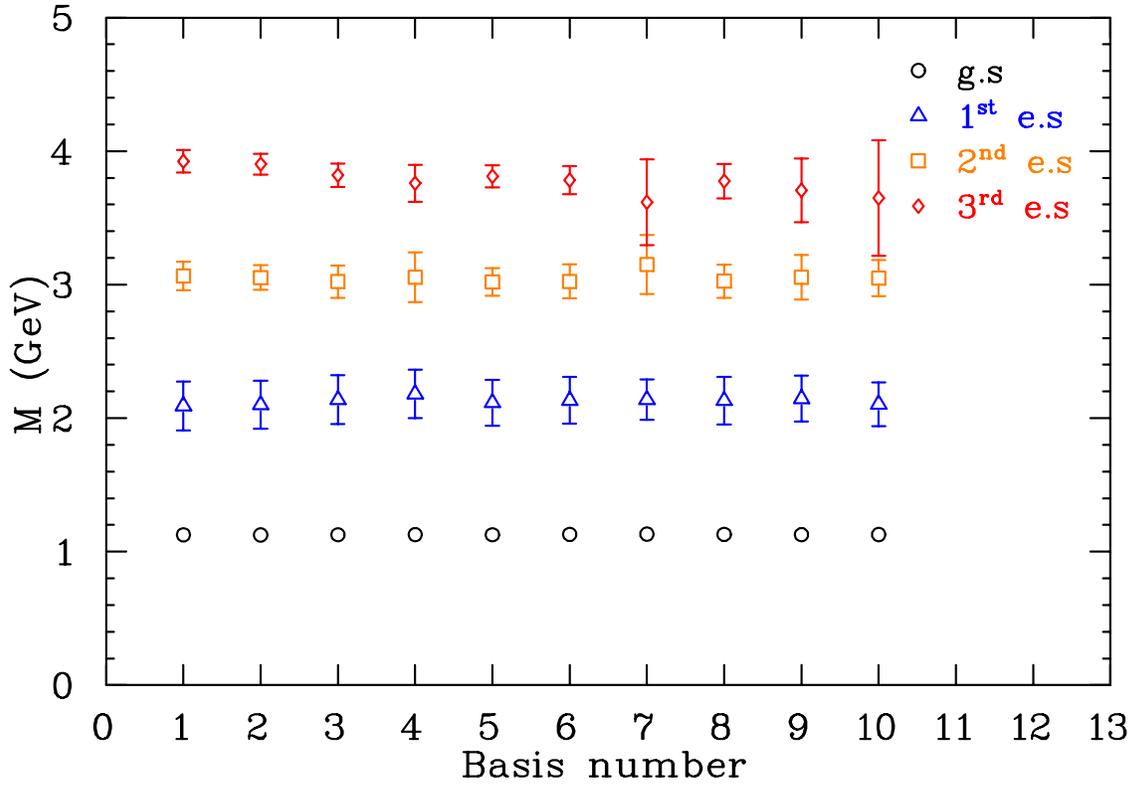} 
  \end{center}
    \caption{(Color online). Masses of the $N{\frac{1}{2}}^{+}$ energy
      states for various $4\times 4$ correlation matrices (bases), given in
      Table~\protect\ref{table:varieties_of_4x4_matrices}, for
      $k_{ud}=0.13770$, over 50 configurations.}
 \label{fig:4x4_x1x1_Kud01377000Ks01364000}
\end{figure*}

Consistency of the extracted masses is manifest in
Fig.~\ref{fig:4x4_x1x1_Kud01377000Ks01364000}. In particular, the ground and
  Roper states are robust. The highest energy state shows some basis
 dependency (smearing dependency) which is to be expected as this
 state must accommodate all remaining spectral strength~\cite{Mahbub:2009aa}.

 It is noted that basis operators that are linearly dependent will cause the
  eigenvalue analysis to fail as there will be a singularity in the
  correlation matrix.  The fact that our analysis succeeds
  indicates that our choice of operators access an equal number of dimensions
  in the Hilbert space. It is interesting to examine the stability of
  the masses to different choices of bases to ascertain whether one
  has  reliably isolated single eigenstates
  of QCD.  The relevant issues are: (i) whether or not the
  operators are sufficiently far from collinear that numerical errors
  do not prevent diagonalisation of the correlation matrix and, (ii)
  whether or not the states of interest have significant overlap with
  the subspace spanned by our chosen sets of
  operators. Since our correlation matrix
  diagonalisation succeeded, except at large Euclidean times where
  statistical errors dominate, we conclude that our operators
  are sufficiently far from collinear.

 Basis numbers 4, 7, 9 and 10 contain
 higher smearing-sweep counts of 400 and 800, which results in a significant
 enhancement of errors for the second and third excited states. It is
 noted that the sources with sweep counts of 400, 800 and 1600 are very
 challenging as the smearing radii for these sources are close to the
 wall source. The  poor signal-to-noise ratio for these sources make the
 correlation  matrix
 analysis more challenging and the eigenvalue analysis becomes
 unsuccessful for a large number of variational parameters
 ($t_{0},\triangle t$). Therefore, the sources 400, 800 and 1600 are undesirable to work
 with.

\begin{figure*}[!h]
\vspace{0.5cm}
  \begin{center}
  \includegraphics[height=0.90\textwidth,angle=90]{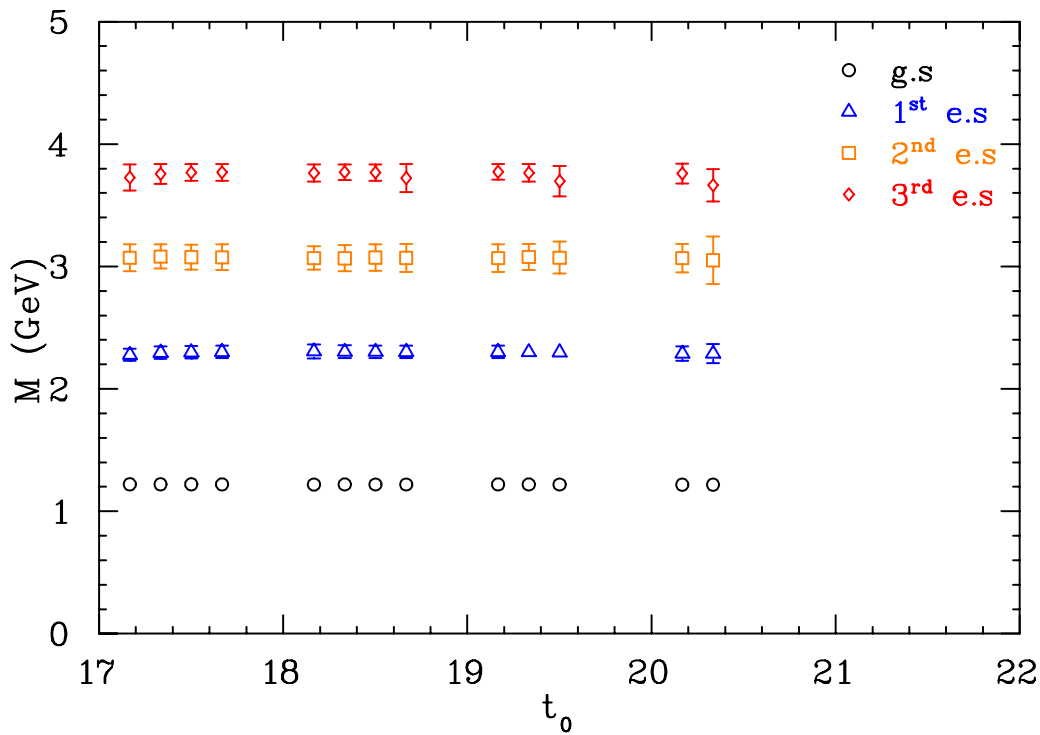} 
 \end{center}
\vspace{-0.5cm}
 \caption{(Color online). Masses of the nucleon, $N{\frac
          {1}{2}}^{+}$ states,  from the projected correlation
        functions as shown in Eq.~(\protect\ref{projected_cf_final}). Each set
        of ground (g.s) and excited (e.s) states masses correspond to the
       diagonalization of the correlation matrix for each set of
       variational  parameters $t_{0}$ (shown in major tick
       marks) and $\triangle t$ (shown in minor tick marks). Figure
   corresponds  to  $k_{ud}=0.13754$ and for the 3rd
   basis. } 
 \label{fig:ma_Kud_013754_4x4_x1x1}
\end{figure*}

The agreement among the three lowest lying eigenstates is remarkable
and verifies that our approach successfully isolates true
eigenstates~\cite{Mahbub:2009aa,Mahbub:2009nr}.
 
Basis number 3 has good diversity
including both lower and higher smearings which is necessary for
 the extraction of masses over the entire heavy to light quark mass range. As
a result, basis number 3 is considered for the following more
extensive  correlation-matrix analysis.

\begin{figure*}[!t]
\vspace{0.5cm}
  \begin{center}
  \includegraphics[height=0.90\textwidth,angle=90]{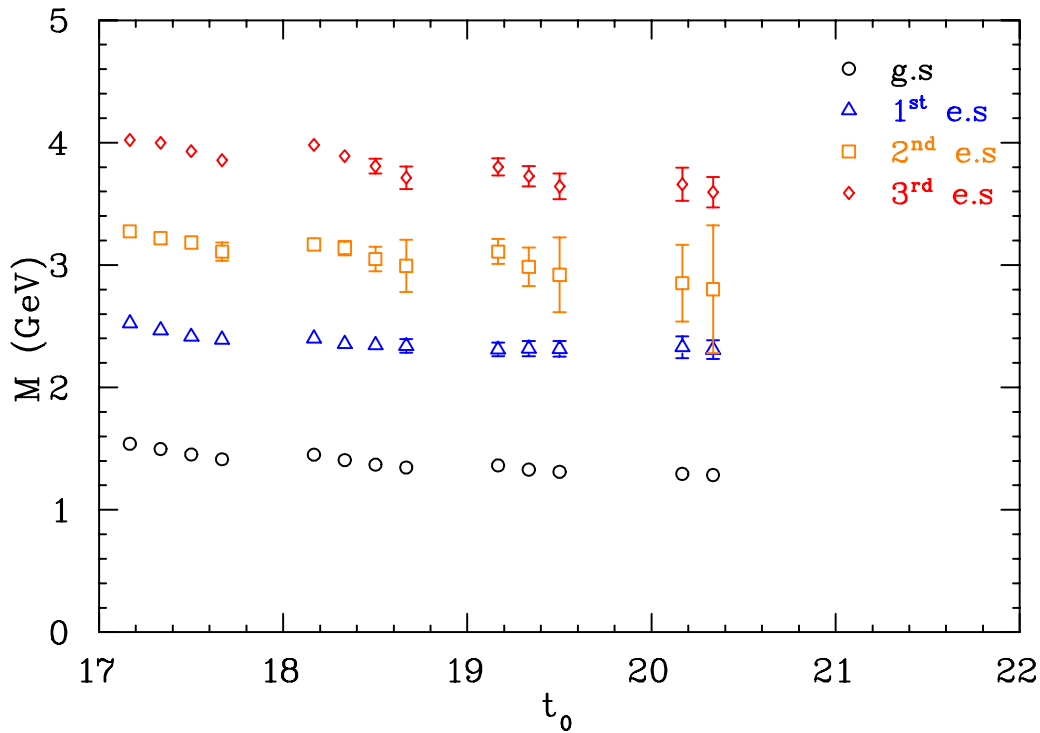} 
 \end{center}
\vspace{-0.5cm}
 \caption{(Color online). As in Fig.~\protect\ref{fig:ma_Kud_013754_4x4_x1x1},
   but masses are calculated from the eigenvalues as shown in
   Eqs.~(\protect\ref{eq:right_evalue_eq}) and~(\protect\ref{eq:left_evalue_eq}).} 
 \label{fig:ei_Kud_013754_4x4_x1x1}
\end{figure*}

Consistency of the extracted masses from projected correlation
functions over the variational parameters $t_{0}$ and $\triangle t$ is
evident  in
Fig.~\ref{fig:ma_Kud_013754_4x4_x1x1}, whereas a significant
dependence of the extracted masses from the eigenvalues is evident in
Fig.~\ref{fig:ei_Kud_013754_4x4_x1x1}. Asymptotically, these energies
agree with those of Fig.~\ref{fig:ma_Kud_013754_4x4_x1x1}. These
aspects of  the extracted masses on variational
parameters is in complete accord with our quenched
analysis presented in Refs.~\cite{Mahbub:2009aa,Mahbub:2010jz,Mahbub:2010me,Mahbub:2009nr}.

\section{Conclusions}

In these proceedings, we have presented a systematic correlation matrix
analysis method using a variety of fermion field smearings at the
source and sink to extract
excited-state energies of the nucleon. Of particular interest is the Roper
resonance from 2+1 flavor QCD. The negative parity channel will also
be investigated to obtain information about the level orderings
between the Roper and ${N\frac{1}{2}}^{-}$ ground state in full
QCD. Performing this analysis technique at all the quark masses will
be the  subject of future investigations.

\section{Acknowledgments}

We thank
PACS-CS Collaboration for making  these 2+1 flavor
configurations available and the ILDG~\cite{ildg:site} for creating the opportunity,
tools and formalism for sharing these configurations. This research
was  undertaken on the NCI National  Facility in Canberra,
Australia, which is supported by the Australian Commonwealth
Government. We also acknowledge eResearch SA for generous
grants of supercomputing time which have enabled this project.  This
research is supported by the Australian Research Council.

\end{document}